\documentstyle[11pt]{article}
\input epsf
\begin{document}
\hsize = 6.0 in
\vsize =11.7 in
\hoffset=0.1 in
\voffset=-0.5 in
\baselineskip=20pt
%\begin{document}
\newcommand{\ghat}{{\hat{g}}}
\newcommand{\Rhat}{{\hat{R}}}
\newcommand{\ih}{{i\over\hbar}}
\newcommand{\Scal}{{\cal S}}
\newcommand{\fudge}{{1\over16\pi G}}
\newcommand{\tn}{\mbox{${\tilde n}$}}
\newcommand{\mg}{\mbox{${m_g}^2$}}
\newcommand{\mf}{\mbox{${m_f}^2$}}
\newcommand{\hk}{\mbox{${\hat K}$}}
\newcommand{\vk}{\mbox{${\vec k}^2$}}
%\renewcommand{\thesection}{\arabic{section}}
%\def\theequation{\thesection.\arabic{equation}}
%\newcommand{\sect}[1]{\section{#1} \setcounter{equation}{0}}
%\newcounter{letter}
\newcommand{\eqletter}{ \hfill (\theequation\alph{letter})}
\newcommand{\gm}{{(\Box+e^2\rho^2)}}
\newcommand{\eql}{\nonumber &\eqletter \cr
                  \addtocounter{letter}{1}}
\newcommand{\beq}{\begin{equation}}
\newcommand{\eeq}{\end{equation}}
\newcommand{\bea}{\begin{eqnarray}}
\newcommand{\eea}{\end{eqnarray}}
\newcommand{\beal}{\setcounter{letter}{1} \begin{eqnarray}}
\newcommand{\eeal}{\addtocounter{equation}{1} \end{eqnarray}}
\newcommand{\none}{\nonumber \\}
\newcommand{\req}[1]{Eq.(\ref{#1})}
\newcommand{\reqs}[1]{Eqs.(\ref{#1})}
\newcommand{\larrow}{\,\,\,\,\hbox to 30pt{\rightarrowfill}
\,\,\,\,}
\newcommand{\slarrow}{\,\,\,\hbox to 20pt{\rightarrowfill}
\,\,\,}
\newcommand{\half}{{1\over2}}
\newcommand{\bfx}{{\vec{x}}}
\newcommand{\bfy}{{\vec{y}}}
\newcommand{\zfp}{Z_{{FP}}}
\newcommand{\zf}{Z_{{F}}}
\newcommand{\zr}{Z_{{R}}}
\newcommand{\zop}{Z_{{OP}}}
\newcommand{\zekt}{Z_{EKT}}
\newcommand{\phstar}{{\varphi^\dagger}}
\newcommand{\ham}{{\cal H}_t}
\newcommand{\diff}{{\cal H}_r}
\begin{center}
{\bf Back Reaction of Hawking Radiation on Black Hole Geometry}\\
\vspace{15 pt}
{\it by}\\
\vspace{13 pt}
H. Zaidi${}^1$ {\it and} J. Gegenberg${}^2$ \\[5pt]
  ${}^1${\it  Department of Physics
}\\
{\it University of New Brunswick}\\
{\it Fredericton, New Brunswick}\\
{\it CANADA, E3B 5A3}\\
{\sl e-mail:  zaidi@unb.ca}\\[5pt]
${}^2${\it Department of Mathematics and Statistics}\\
   {\it University of New Brunswick}\\
   {\it Fredericton, New Brunswick}\\
   {\it CANADA E3B 5A3}\\
{\sl e-mail:  lenin@math.unb.ca}
\end{center}
\vspace{40pt}
{\narrower\smallskip\noindent
{\bf Abstract} :
We propose a model for the geometry of a {\it dynamical} spherical
shell in which the metric is asymptotically Schwarzschild, but
deviates from Ricci-flatness in a finite neighbourhood of the shell.
Hence, the geometry corresponds to a `hairy' black hole, with the 
hair originating on the shell.  
The metric is regular for an infalling shell, but it bifurcates,
leading to two disconnected Schwarzschild-like 
spacetime geometries.  The shell is interpreted as either collapsing
matter or as Hawking radiation, depending on whether or not the shell
is infalling or outgoing.  In this model, the Hawking radiation 
results from tunnelling between the two geometries.  
Using this model, the back reaction
correction from Hawking radiation is calculated.}

\vspace{40 pt}

\bigskip\noindent
UNB Technical Report 97-02\\

\bigskip
\begin{center}
{\it May, 1997}
\end{center}
\clearpage
%%%%%%%%%%%%%%%%%%%%%%%%%%%%%%%%%%%%%%%%%%%%%%%%%%%%%%%%%%%%%%%%%
\section{Introduction}
The back reaction of Hawking radiation on the black hole geometry has
attracted considerable attention during recent years \cite{kw}.
However, explicit calculation of the resulting corrections to the
spectrum have only been performed in the semi-classical approximation
using a thin shell model for the Hawking radiation field \cite{kw},
wherein a ``dressed shell" wave function was obtained in a reduced
phase-space quantization formalism, with the background metric fixed
to be Ricci-flat, and hence Schwarzschild, everywhere except at the
location of the shell, assumed to be an imbedded 3-manifold.  To
lowest order, the shell traverses a geodesic in the ambient
spacetime.  Near the Schwarzschild radius (given in terms of the ADM
mass $M_A$ by $r=2M_A$), the affine parameter $t$ of the geodesic
acquires a small imaginary part for an outgoing shell.  The Hawking
radiation is thereby generated in a narrow region near the
Schwarzschild radius.  In fact, the spectrum of Hawking radiation is
completely determined by Im\,$t$, independently of the detailed form
of the wave function.

In this paper we examine the dependence of the correction to the spectrum of 
Hawking radiation on the choice of spacetime metric used to model the 
back reaction.  For this purpose, we introduce a more detailed model, 
so that the minimal modification to the Schwarzschild geometry 
reproduces the expected stress-energy tensor in the asymptotic region.  
The possible choices are severly constrained by this asymptotic 
condition as well as the boundary conditions at the shell.  This is 
reviewed in Section 2.  In Section 3, such a metric is displayed 
and its properties are discussed.  In the absence of the back reaction, 
it reduces to the Schwarzschild metric, with the shell playing the role 
of collapsing matter, if it is infalling, and of the Hawking radiation, if 
it is outgoing.  In the next approximation, the results of Kraus and 
Wilczek \cite{kw} are recovered, but with an additional correction of the 
same order of magnitude.  This additional correction arises from the details 
of the dynamics.  We also derive the corrections arising from a non-zero 
shell mass- i.e. from a {\it massive} scalar field.  In 
Section 4, we examine the stress-energy tensor.  In the last section, 
we conclude our discussion and anticipate a more complete quantum 
mechanical treatment of the problem.   

Our model is based on a reparameterization, $r\to R(r;\alpha(\hat
r),\beta(\hat r))$, where $\alpha(\hat r),\beta((\hat r)$ are
parameters depending on the shell position $\hat r$.   The parameters
$\alpha(\hat r), \beta(\hat r)$ are uniquely determined by the
boundary conditions at the shell.  The ADM mass $M_A$ is then
determined from the saddle point condition.  The parameter
$\beta(\hat r)$ satisfies a cubic equation, the physical roots of
which represent two disconnected spacetime geometries.  For an
outgoing shell, we find that $\beta(\hat r)$ acquires a small
imaginary part in the narrow {\it non-classical} region around the
Schwarzschild radius.  It is in this region that the Hawking
radiation originates from tunnelling bwtween the two spacetime
geometries.  We obtain an expression for Im\,$t$ from the solution of
the geodesic equation- the lowest order approximation to the motion
of the shell.  A notable feature of this approach is that the
geodesics for an infalling particle, as a function of Schwarzschild
time $t$, are complete.  This is due to the dynamical distortion of
the Schwarzschild metric- the back reaction.  Hence, the affine
parameter of these geodesics can be used as the time variable in the
calculation of the Hawking spectrum.

\section{Review of Spherical Shells and Hawking Radiation}

We will consider a spherically symmetric geometry in which there is a 
thin spherical shell with mass $m$.  In terms of spherical coordinates 
$(t,r,\theta,\phi)$, the metric is of the form
\beq
ds^2=-N^2dt^2+L^2\left(dr+N^rdt\right)^2+R^2\left(d\theta^2+\sin^2{\theta}\,
d\phi^2\right),\label{4metric}
\eeq
where $N,L,N^r,R$ are functions of $(t,r)$ only.  In fact, the dependence on 
the time coordinate $t$ is only through $\hat r(t)$, where $r=\hat r(t)$ 
describes the position of the spherical shell.  

The action can be written in ADM form \cite{fischler,kw}:
\beq
S[\hat r,R,L,N,N^r]=\int dt\left[\hat p\dot{\hat r}-M_A+\int dr\left(\Pi_R \dot R 
+\Pi_L \dot L-N\ham-N^r\diff\right)\right].\label{action}
\eeq
In the above, $\hat p,\Pi_R,\Pi_L$ are the canonical momenta conjugate, 
respectively, to $\hat r,R,L$.  The term $M_A$ is a boundary term which is 
a constant of the motion and may be identified with the ADM mass.  We 
use units in which $G=c=1$.  The 
lapse $N$ and shift $N^r$ functions are Lagrange multipliers enforcing the 
constraints:
\bea
\ham&:=&{L\Pi_L^2\over2 R^2}-{\Pi_L\Pi_R\over R}+\left({RR'\over L}\right)'
-{(R')^2\over2L}-{L\over2}+{\hat E\over\hat L}\delta(r-\hat r(t))\approx 0,\\
\diff&:=&R'\Pi_R-L\Pi_L'-\hat p\delta(r-\hat r(t))\approx 0.\label{constraints}
\eea
These are respectively the Hamiltonian and diffeomorphism constraints.  Hatted 
quantities, e.g. $\hat L$ are given by $\hat L(t)={L(r,t)}_{\displaystyle \mid_{r=\hat r(t)}}$.   Also $\dot L=\partial L/\partial t, L'=\partial L/\partial r$, etc.  
Finally, the quantity $\hat E:=\left(\hat p^2+m^2\hat L^2\right)^\half$.

The constraints can be solved for the field momenta \cite{fischler,geg}
\beq
\Pi_R= L\Pi_L'/R',\mbox{if $r\neq \hat r(t)$};\label{PiR}
\eeq
\beq
\Pi_L=\left\{\begin{array}{ll}
R\left[(R'/L)^2-1+2M_A/R\right]^\half, & \mbox{if $r>\hat r(t)$}\\[2mm]
R\left[(R'/L)^2-1+2M/R\right]^\half, & \mbox{if $r<\hat r(t)$}
\end{array}\right.\label{PiL}
\eeq
Moreover, the discontinuity at the shell gives the following boundary conditions:
\beq
\Delta\Pi_L=-\hat p/\hat L;\,\,\, \Delta R'=-\hat E/\hat R,\label{junction}
\eeq
where 
$$
\Delta\Pi_L:=\lim_{\epsilon\to 0^+}\left(\Pi_L(\hat r(t)+\epsilon) - 
\Pi_L(\hat r(t)-\epsilon)\right),
$$ 
etc.  As usual in thin shell formulations, it is assumed that the
metric is continuous, but the first deriviatives normal to the shell are 
discontinuous at $r=\hat r(t)$.

A semi-classical quantization of the system can be performed either 
in terms of the reduced phase space \cite{kw}, or via Dirac quantization, 
wherein the constraints are imposed as operators which annihilate physical 
states \cite{fischler,geg}.  In either case, one ends up with a second-
quantized  
wave function $\psi(t,r)$ associated with the location of the shell.  
When expanded in modes the wave-function has the form:
\beq
\psi(t,r)=\int{d\omega\over2\pi}\left[a_i(\omega)u_i(\omega,t,r)+
a^\dagger_i(\omega)u^*_i(\omega,t,r)\right],
\eeq
where the index $i=1,2$ is not summed over, and denotes, respectively the 
quantities defined with respect to observers falling with the shell and 
in the asymptotic region.  In particular, $a_i(\omega), a^\dagger_i(\omega)$ 
are annihilation and creation operators, while the $u_i$ and their 
complex conjugates $u^*_i$ are the mode functions.  The mode functions 
$u_2$, in the asymptotic region, are given by $u_2=u(\omega,r)
e^{-i\omega t}$.  The operators corresponding to the different observers 
are related by Bogoliubov transformations, with the coefficients $A(\omega, 
\omega'), B(\omega,\omega')$ given by the 
Fourier transforms of the wave-function:
\bea
A={1\over2\pi u(\omega,r)}\int^\infty_{-\infty}dt\, e^{i\omega t}\psi(t,r),\\ 
B={1\over2
\pi u^*(\omega,r)} 
\int^\infty_{-\infty}dt\, e^{i\omega t}\psi^*(t,r).\label{bogol}
\eea
In the saddle point limit one gets the result \cite{kw}  
$M_A=M\pm\omega$, where the $+(-)$ sign refers to an ingoing (outgoing) shell.  
This fixes $M_A$ in the semi-classical approximation.\cite{ff1}

By standard arguments \cite{hh,bd,kw}, one may compute 
the number of outgoing particles in Hawking radiation with frequencies 
between $\omega$ and $\omega+d\omega$ :
\beq
n(\omega){d\omega\over2\pi}=\left[{\left|A/B\right|}^2-1\right]^{-1}{d\omega
\over2\pi}.\label{spec}
\eeq
The quantity $n(\omega)$ gives the spectrum if the contribution of 
those particles falling back 
into the black hole can be ignored.  
If back reaction is also neglected, 
$n(\omega)$ is given by the well-known thermal spectrum
\cite{bd,hh}
\beq
n(\omega)={1\over e^{8\pi M\omega}-1}.\label{spectrum}
\eeq

\section{The Back Reaction}

In the absence of the shell, the one-parameter family of
Schwarzschild metrics is the unique solution of the vacuum Einstein
equations with spherical symmetry.  The presence of the shell
disrupts this uniqueness.  In fact, classically the metric would be
Schwarzschild everywhere except at the location of the shell itself.
The metric has a jump discontinuity at the location of the shell, and
the Schwarzschild metrics on either side of the shell can have
unequal Schwarzschild mass.  In a quantum mechanical treatment,
classical geometry may not be completely established.  In a
semi-classical analysis, there is a transitory regime of duration
$\Delta\tau$ specified by the quantum uncertainty principle, during
which the Schwarzschild mass, or total energy, $M$ is uncertain by
$\Delta M$, where $\Delta M\Delta\tau\sim\hbar$.  We expect that in
this approximation, the uncertainty in total energy is approximately
given by the energy of the shell, $\Delta M\sim\hbar\omega$.\cite{ff1}  

We can model the situation via the ``semi-classical Einstein equations"
\beq
G_{\mu\nu}(x,\omega)=8\pi T_{\mu\nu}(x,\omega),\label{semi}
\eeq
where the stress-energy tensor $T_{\mu\nu}(x,\omega)$ includes the Hawking 
radiation itself, with a concommitant distortion of the ``background" 
Schwarzschild 
geometry.  The program would begin with consideration of some matter field, 
say a scalar field governed classically by the Klein-Gordon equations, 
propogating in the distorted Schwarzschild background.  However, in the absence of 
an exact classical solution which describes the back reaction effect,   
we will propose a physically reasonable metric as a ``solution" of \req{semi}.  
For this purpose, we will rely on the constraints and the boundary conditions 
at the shell to limit the possibilities.  In addition to this, the 
stress-energy tensor $T_{\mu\nu}$ will be required to be consistent with 
well-known predictions for the flux of Hawking radiation in the asymptotic 
region, i.e., as $\hat r\to\infty$.  Moreover, Hawking radiation may be 
viewed as arising from a mismatch between the reference frames of an
observer at infinity and that of an infalling observer \cite{hh}.  This 
mismatch is represented by the Bogolubov coefficients given by \req{bogol}.  
Therefore, in our model, the distorted metric must approach that of an 
infalling observer as $r\to\hat r(t)$.  We can satisfy this requirement by 
choosing coordinates such that in the ``inner region" $r<\hat r(t),\, 
\Pi_L=0=\Pi_R$.  These conditions severly limit the choice of metric.

We have found the following metric satisfies the above requirements:
\beq
R=\left\{\begin{array}{ll} R_+:=r\left[1+\gamma(r,\hat r(t),\omega)\right],& 
\mbox{if $r>\hat r(t)$}\\[2mm]
R_-:=r\left[1+\beta(\hat r(t),\omega)\right],&\mbox{if $r<\hat r(t)$}
\end{array}\right.\label{metricR}
\eeq
\beq
L=\left\{\begin{array}{ll}
L_+:=\left(1-2M_A/R_+\right)^{-\half},&\mbox{if $r>\hat r(t)$}\\
L_-:=\left(1+\beta(\hat r(t),\omega)\right)\left(1-2M/R_-\right)^{-\half}
,&\mbox{if $r<\hat r(t)$}
\end{array}\right.\label{metricL}
\eeq
Furthermore, we will choose coordinates such that the lapse and 
shift functions are $N=1/L,\,N^r=0$. 

The functional dependence of the parameters 
$\gamma(r,\hat r,\omega),\beta(\hat r,\omega)$ on $\hat r$ will be determined 
by the boundary conditions at the shell and in the asymptotic region.  

In the asymptotic region, given by $r\to \infty$, we require that $R_+\to 
r$.  Hence 
\beq
\gamma(r,\hat r,\omega)\to 0,
\eeq
in the above limit.  This ensures that the metric 
is asymptotically Schwarzschild. 

The continuity of $R$ at the shell requires that $\gamma(\hat r,\hat r,\omega) 
=\beta(\hat r,\omega)$.  
Furthermore, the derivative of $\gamma(r,\hat r,\omega)$ with 
respect to its first 
argument, evaluated at $r=\hat r$ is determined from the junction conditions 
\req{junction} and the constraints \req{PiR},\req{PiL}.  The result is:
\beq
\hat r\alpha={(2+\beta)\over 2(1+\beta) }+{\left(m/\hat r\right)^2\over 
2\beta\left(1+\beta-2M/\hat r\right)},
\label{alpha}
\eeq
where we have written:
\beq
\alpha(\hat r,\omega):={-1\over\beta(\hat r,\omega)}{\partial\over\partial r}
{\gamma(r,\hat r,\omega)}_{\displaystyle \mid_{r=\hat r}}.\label{solvalpha}
\eeq
We specify the function $\gamma(r,\hat r,\omega)$ only up to the correct 
limiting behaviour, as discussed above.

It remains to ensure that $L$ is continuous across $r=\hat r(t)$.  This will 
determine the parameter $\beta$.  The equation $L_+=L_-$ at $r=\hat r(t)$ 
implies that
\beq
x^3-(hf)x^2-x+f=0,\label{cubic}
\eeq
where 
\beq
x:=1+\beta;\,\,\,f:=2M/\hat r;\,\,\,h:=M_A/M.
\eeq
The three roots of this cubic are given by
\beq
P_++P_-+{hf\over3};\,\,\,\,\,\,-{P_++P_-\over2}\pm i\sqrt{3}\left({P_+-P_-\over2}\right)
+{hf\over3},\label{roots}
\eeq
where
\bea
a=&-1-{1\over3}(hf)^2;\,\,\,b=-{2\over27}(hf)^3-{1\over3}(hf)+f;\\
c=&\displaystyle \sqrt{{b^2\over4}+{a^3\over27}};\\
P_\pm=&\displaystyle \left(-{b\over2}\pm c\right)^{1/3}.
\eea
If $c^2>0$, there is one real and a complex conjugate pair of roots;  if 
$c^2<0$, there are three real roots.

The nature of the roots depends on the value of $h$.  In fact,  
$h=1\pm\omega/M$, with the ($+$) sign for ingoing, and the ($-$) sign for 
outgoing radiation.  It is easy to see that all the roots are real
for ingoing radiation ($h>1$), but complex roots are possible for
outgoing radiation.  Figures 1a and 1b display the behaviour of the roots, 
for infalling and outgoing shells, respectively.\cite{ff2}

\begin{figure}[h]
\epsfxsize 6.0in
\begin{center}
\epsfbox{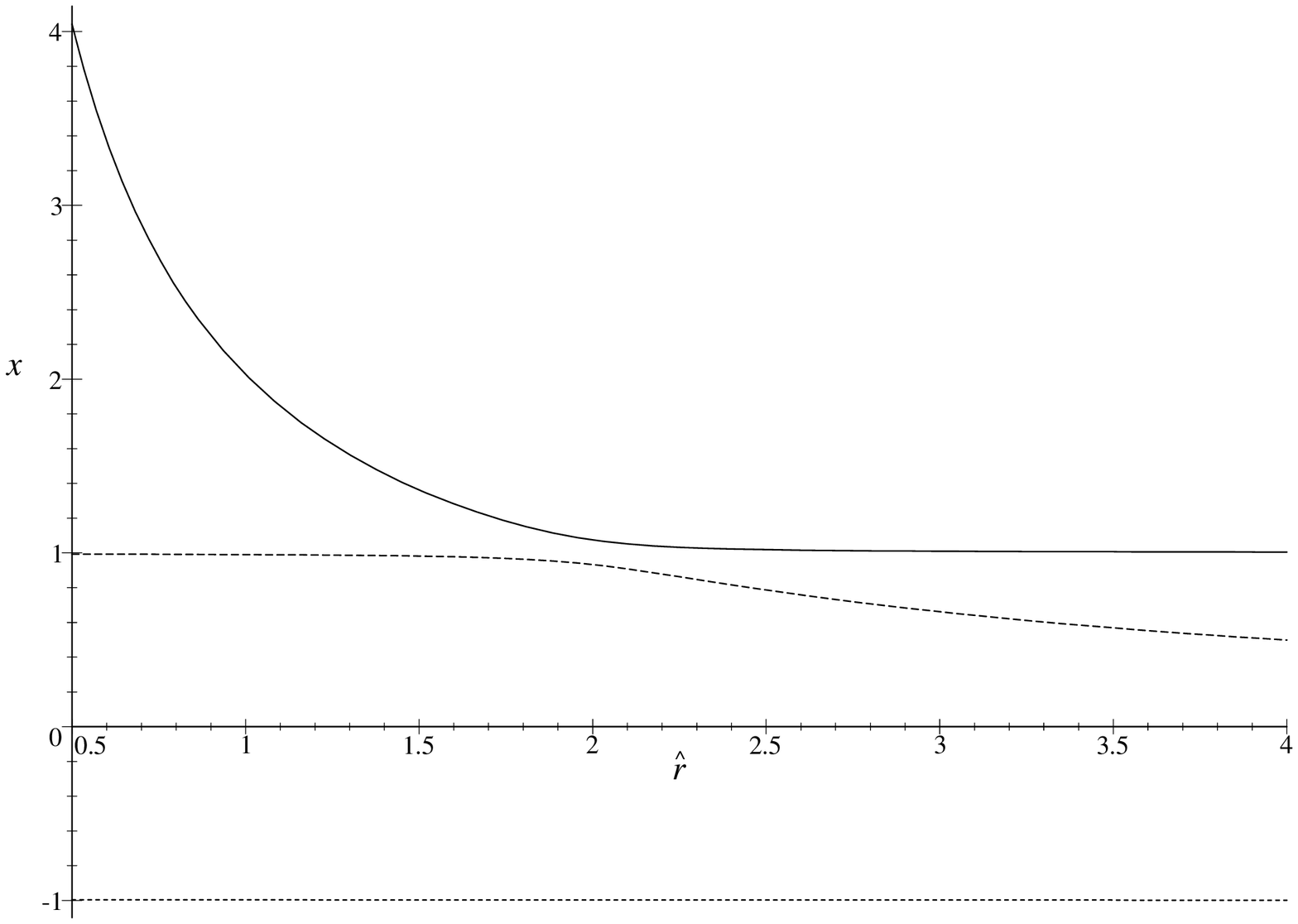}

Figure 1a: $x$ vs. Shell Position $\hat{r}$, $h=1.01$\hspace*{20mm}
\end{center}
\end{figure}
\begin{figure}[h]
\epsfxsize 6.0in
\begin{center}
\epsfbox{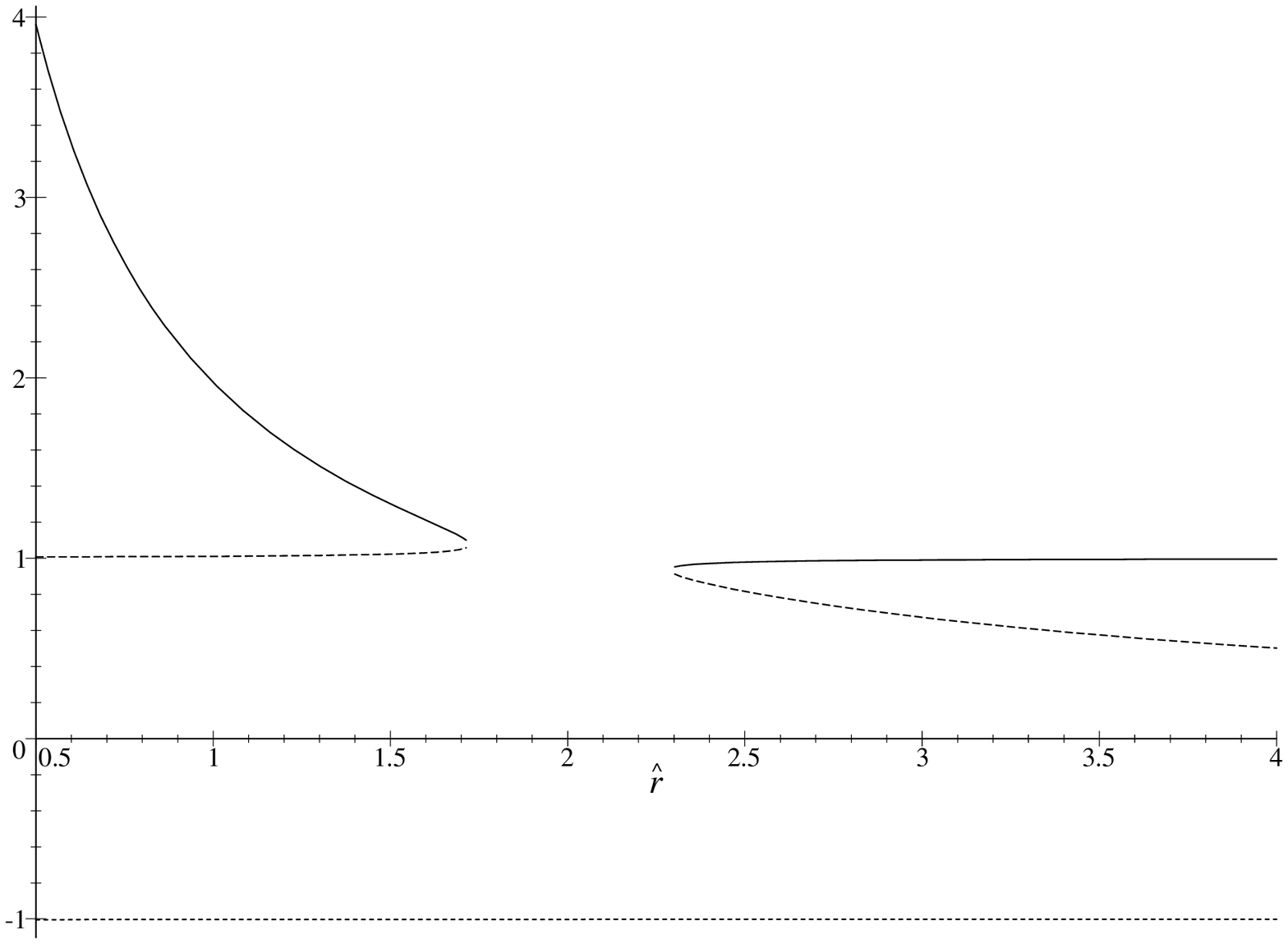}

Figure 1b: $x$ vs. Shell Position $\hat{r}$, $h=0.99$\hspace*{20mm}
\end{center}
\end{figure}
The behaviour of the roots in the respective regions where  
$\hat r>>2M$ or where $\hat r\approx 2M$ is most important.  In the former, 
where the shell is far from the horizon, $x$ is very close to 1.  We call 
this root ``physical", since for it, the metric is asymptotically flat.  
We note that for the physical root, $\beta=\pm\omega/\hat r +O\left(\hat r^{-2}
\right)$ for the $+$ ($-$) sign for ingoing (outgoing) shell.

In the other region, near the event horizon, we expand in powers of the 
two parameters $\omega/M$ and $\epsilon:=\hat r-2M$.  To lowest order, for the 
outgoing shell, the roots are
\beq
-1-{9\over4}\left({\omega\over M}\right);\,\,\,\,\,\,1-{1\over8}\left({\omega
\over M}\right)-\half\left({\epsilon\over2M}\right)\pm i\Delta,\label{rootsh}
\eeq
where
\beq
\Delta:=\sqrt{\omega\over2M}\left[1-{5\over8}\left({\omega\over M}\right)
+\half\left({\epsilon\over2M}\right)\right].
\eeq

To summarize, the physical root $1+\beta$ of \req{cubic} is always real 
for an ingoing shell, but for an outgoing shell, it acquires a small 
imaginary part in a narrow region around $\hat r=2M$.  This is the region 
where Hawking radiation originates and escapes by way of tunnelling 
through the classical barrier.  In this region the solution of the geodesic 
equation can become complex, as will be shown below.

At the classical level, each point on the spherical shell traverses a 
geodesic.  The geodesic equation can be obtained directly from the 
boundary condition \req{junction} and from the expression for the momentum 
canonically conjugate to $\hat r$, 
\beq
\hat p=m\hat L^2\dot{\hat r}\left[\hat N^2-\hat L^2\dot{\hat r}^2\right]^{-\half}.
\eeq
The result is that
\beq
\dot{\hat r}=\pm\hat N\left(\Delta\Pi_L/\hat R\right)/\Delta R',
\label{geo}\eeq
where the $+$ ($-$) sign applies to an outgoing (ingoing) geodesic.  
(If $m=0$, it follows from \req{geo} and from \req{junction} that 
$\dot{\hat r}=\pm \hat N$, so that the points on the shell traverse 
null geodesics.)

The geodesic equation \req{geo} is integrated to yield
\beq
t=-\left\{\pm\int{d\hat r\hat R\over \hat R-2M_A}\,{1\over2(1+\beta)\sqrt
{\left(\hat R_+'\right)^2-1}}\left[\beta(2+\beta)+{\hat R\over \hat R-
2M}\left({m\over\hat r}\right)^2\right]\right\}\,\,\,+t_0,\label{geosol}
\eeq
where $\hat R=\hat r(1+\beta)$ and $r_0=$constant.  For $\beta=0$ and 
$m=0$, this 
reduces to the usual equation for a null geodesic in Schwarzschild spacetime:
\beq
t=\pm\left[\hat r+2M\ln{(\hat r/2M-1)}\right]+t_0.
\eeq
However, if $\beta$ is complex, then $t$ becomes complex.  In this case, 
the integral has a branch cut along the real axis in the 
complex $\hat r$-plane.  The region in which $\beta$ is complex is quite 
narrow, centered around $\hat r=2M_A$.  In this region, $\beta\approx$ 
constant, and the branch cut can be replaced with a simple pole. The 
imaginary part of $t$ can then be evaluated as the residue at the pole, with 
the result:
\beq
t=t_{\mbox{real}}+i4\pi M_A\left\{1-{3\over8}\left({\omega\over M_A}\right)
\left[1+{1\over16}\left({m\over M_A}\right)^2\right]\right\}.\label{time}
\eeq
Here we have also assumed that $m<<\omega$.  Only the lowest order 
corrections in $\omega/M_A,m/M_A$ and $m/\omega$ are retained \req{time}.  
If $m=0$, one obtains
\beq
\mbox{Im}\,t=4\pi M_A(1-3\omega/8M_A)=4\pi(M_A-3\omega/8).\label{Imt}
\eeq 
If the last term is ignored, we have the result of Ref.\ 1:  $\mbox{Im}\,t=
4\pi M_A=4\pi (M-\omega)$.  If all the corrections of order $\omega/M$ are 
ignored, we have Hawking's result:  $\mbox{Im}\,t=4\pi M$, obtained in the 
Euclidean formulation.\cite{bd,hh}

The spectrum of Hawking radiation is given by \req{bogol} and \req{spec}.  
By virtue of \req{time} we obtain:
\beq
[n(\omega)]^{-1}=\exp\left\{8\pi\omega\left[M_A-{3\over8}\omega\left(
1+{m^2\over16M_A^2}\right)\right]\right\}-1.
\label{3.23}
\eeq
For the case of a massless shell ($m=0$), we have:
\beq
[n(\omega)]^{-1}=\exp\left[8\pi\omega\left(M_A-3\omega/8\right)\right]-1. \label{3.24}
\eeq
With $M_A=M-\omega$, we recover the results of Ref.\ 1, if the additional 
correction of $3\omega/8$ is ignored.  The origin of this additional 
correction lies in the details of the dynamics, which are sensitive 
to the form of the metric.  Our choice of metric, given by 
\req{metricR} and \req{metricL}, 
reproduces the results of Ref.\ 1, if the details of the shell dynamics in 
the distorted background, represented by \req{geosol} are ignored.  However, 
this does not seem to be justified since the additional correction in 
\req{3.23} is of the same order in $\omega$ as the corrections in Ref.\ 1.  
It appears that the back reaction correction is sensitive to the details of the 
metric.  The ambiguity in the metric can only be diminished by an exact 
semi-classical solution of the Einstein equations.  In the absence of such 
solutions, we have to rely on the consistency of our choice of metric.  We 
will confirm this consistency through a calculation of the stress-energy 
tensor $T_{\mu\nu}$ in the next section.

\section{The Stress-Energy Tensor}
We can calculate $T_{\mu\nu}$ using the standard thin shell formalism \cite
{shell}.  We use Gaussian normal coordinates $(\eta,x^i,i=1,2,3)$ so that 
the metric is of the form:
\beq
ds^2=d\eta^2+h_{ij}dx^idx^j,
\eeq
where $\eta(x^i)=0$ is the equation of the shell and $h_{ij}$ is the 
induced metric on the hypersurface $\Sigma_r$ swept out by the trajectory of the shell.  
In our case, $\eta=r-\hat r(t)$ for a spherical shell.  The stress-energy 
tensor takes the form \cite{shell}
\beq
T_{\mu\nu}=\delta(\eta)\, S_{\mu\nu}+\theta(\eta)\,T^+_{\mu\nu}+\theta(-\eta)
\, T^-_{\mu\nu},\label{4.2}
\eeq
where $\theta$ is the unit step-function.  The tensor $S_{\mu\nu}$ is the 
surface stress-energy carried by the shell, while $T^\pm_{\mu\nu}$ denote the 
regular background contributions on the two sides of the shell.  
Hence, $T^\pm_{\mu\nu}$ describe the black hole hair originating 
on the shell.  Both fall off rapidly with distance from the shell.  Therefore 
Hawking radiation contributes to $S_{\mu\nu}$, but not directly to $T^\pm_
{\mu\nu}$.  The latter represent the transient stress-energy originating 
in the quantum fluctuations of the geometry.  

The surface stress-energy tensor $S_{\mu\nu}$ is determined by the 
extrinsic curvature (second fundamental form) $K_{\mu\nu}$ of $\Sigma_r$ by
\beq
K_{\mu\nu}=\half{\cal L}_n h_{\mu\nu};\,\,\, h_{\mu\nu}:=g_{\mu\nu}-n_\mu
n_\nu,\label{4.3}
\eeq
where $n_\mu$ is the unit normal to $\Sigma_r$.  The Einstein equations now 
take the form:
\bea
S_{ij}&=&-\left(1/8\pi\right) \left(\kappa_{ij}-\kappa h_{ij}\right),
\label{4.4}\\
D_j S^{ij}&=&-T^i_\eta,\label{4.5}\\
\half\left(\hat K^+_{ij}+\hat K^-_{ij}\right)S^{ij}&=&T^\eta_\eta,\label{4.6}
\eea
where $\kappa_{ij}:=\hat K^+_{ij}-\hat K^-_{ij}$, $\kappa:=h^{ij}\kappa_{ij}$ 
and $D_j$ is the covariant derivative with respect to the metric $h_{ij}$.

We are primarily interested in the component $S_{tt}$ for an outgoing 
shell, which determines the surface energy per unit area of the shell.  
Other components of the surface stress-energy tensor determine the 
pressure and tension.  From \req{4.3}--\req{4.4} we obtain for a massless 
shell:
\bea
S_{tt}&=-\hat N^3\hat{\Delta R}'/4\pi\hat R,\\
      &\displaystyle \to -{\omega\over4\pi\hat r^2},\label{4.7}
\eea
as $\hat r\to\infty$.  Therefore, energy {\it outflow} of  Hawking radiation 
from a sphere of radius $\hat r$ at infinity is $\hbar\omega$, as 
expected.  As a result, \req{metricR}--\req{metricL} are consistent with 
the expected asymptotic behaviour of the stress-energy tensor.  Moreover, 
it can be shown that the $T^\pm_{\mu\nu}$ fall off rapidly away from $\Sigma_r
$ and therefore do not contribute to the Hawking radiation.

The total flux of Hawking radiation can now be calculated using
\req{spectrum}.  The total flux is $\int^\infty_0(2\pi)^{-1}d\omega\,
n(\omega)\omega=1/768\pi M^2$, which recovers a well-known result
\cite{bd}.

\section{Analysis of the Geometry}
As we have seen, the spacetime geometry is determined by the
properties of the function $\beta(\hat r,\omega)$.  This function
must satisfy the cubic equation \req{cubic}.  The properties of these
roots are displayed in the graphs of the `optical scalar' $\hat R$
vs. $\hat r$, Figure 2a and Figure 2b.

\begin{figure}[h]
\epsfxsize 6.0in
\begin{center}
\epsfbox{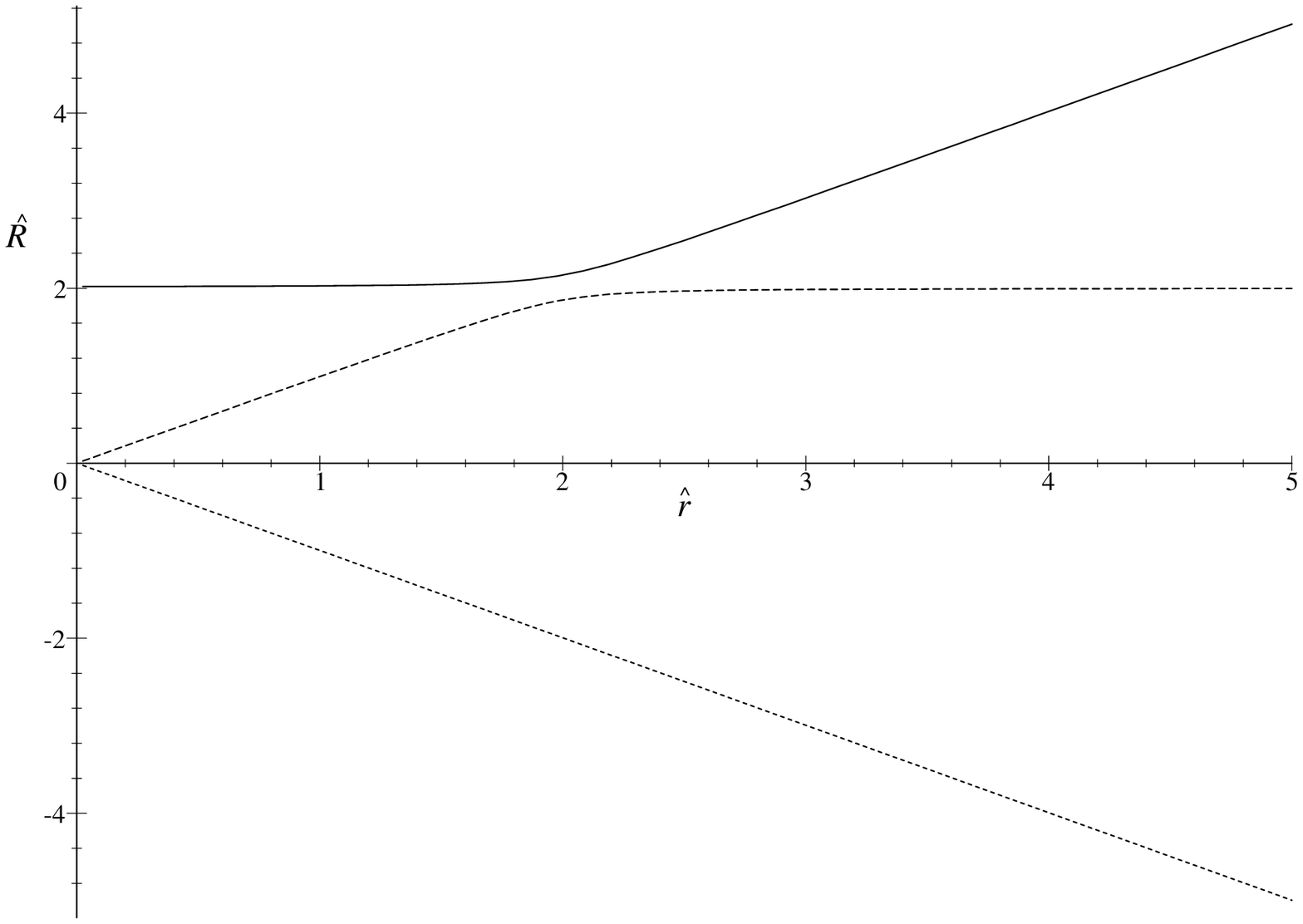}

Figure 2a: Optical Scalar $\hat{R}$ vs. Shell Position $\hat{r}$, $h=1.01$\hspace*{20mm}
\end{center}
\end{figure}
\begin{figure}[h]
\epsfxsize 6.0in
\begin{center}
\epsfbox{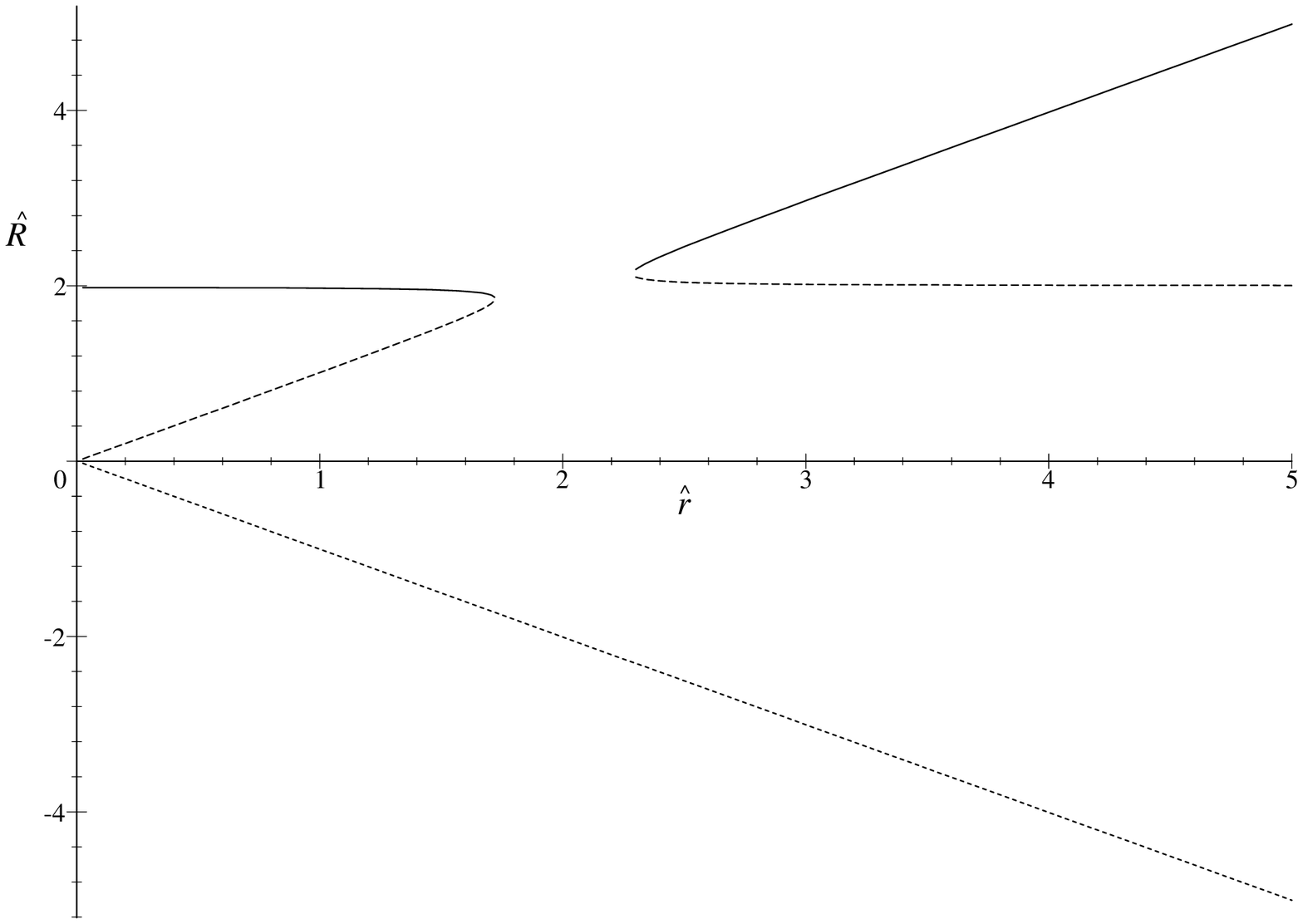}

Figure 2b: Optical Scalar $\hat{R}$ vs. Shell Position $\hat{r}$, $h=0.99$\hspace*{20mm}
\end{center}
\end{figure}
We may conclude from these that the root for which $\hat R< 0$ for all 
$\hat r>0$ 
is unphysical, while the other two roots of the cubic represent a
bifurcation of the spacetime geometry.

The collapsing shell, which has $M_A=M+\omega$, for which the optical
scalar corresponds to a Schwarzschild geometry asymptotically, i.e.
which behaves as $R\sim r$ for large $r$, never reaches the location
$\hat R=2M_A$ except in the limit as $\hat r\to 0$.  The latter is a
singular event in the geometry.  The events at $\hat R=2M_A$ and at
$\hat R=2M$  constitute apparent horizons.  The two geometries are
separated by a gap $2M\leq R\leq 2M_A$.  A collapsing shell inside
the $\hat R=2M$ horizon does fall into the singularity in a finite
time.  However, it cannot cross the apparent horizon $\hat R=2M$ in a
finite time.  Classically these two configurations are distinct
solutions, and represent different spacetime geometries.  It appears
that the shell will only collapse into a black hole via quantum
fluctuations of the spacetime geometry.   Similarly, a shell in the
inner region geometry can enter the outer region only via quantum
tunnelling.   
Since the magnitude of the distortion of the outer geometry 
from that of Schwarzschild is small in the region where
$\hat r$ is near the value $2M_A$; and since, moreover,  $t$ is continuous at
$\hat r=2M_A$, $t$ can be used as an affine parameter.

When $M_A=M-\omega$, i.e., the shell is outgoing and the 
metric is complex in a small region near where the shell is close to 
the horizon.\cite{ff1}  A classical spacetime geometry does not exist in this 
region, and the Hawking radiation from classically 
forbidden configurations takes place through quantum tunnelling.

To summarize, in our model the fixed Schwarzschild  geometry is split
by the back reaction into two distinct spacetime geometries, both of
which contain apparent horizons.  The exterior  geometry approaches
the schwarzschild limit as $\hat r\to\infty$, while the interior
geometry is Schwarzschildean in the limit $\hat r\to 0$.

\section{Conclusion}
We have calculated the back reaction correction for a massless as well as 
a massive thin shell in the semi-classical approximation.  We have found 
an additional correction due to the back reaction for a massless 
shell, so that the total 
correction to $M$ in the Hawking spectrum is $-11\omega/8$ as opposed to 
$-\omega$, obtained in Ref.\ 1.  The correction appears to be sensitive to the 
details of the dynamics, and, therefore, to the choice of metric 
{\it distorted} from that of Schwarzschild by the presence of the shell.  
Our choice satisfies the asymptotic conditions as well as other 
consistencey conditions at the shell.  However, the ambiguity in the 
choice of metric will remain until an exact solution of the semi-classical 
Einstein equations can be found. \cite{ff3}

\bigskip\noindent {\bf Acknowledgments}\\
\noindent
We would like to thank S. Braham, V. Frolov and G. Kunstatter for very 
useful discussions.  We would also like to thank R.G. McKellar for assistance 
with the figures.


\begin{thebibliography}{99}

\bibitem{kw} For a discussion and references, see P. Kraus, {\it Non-Thermal Aspects of Black Hole 
Radiance}, Ph.D. Thesis, Princeton University, 1995 (gr-qc/9508007); P. Kraus and F. Wilczek, Nuc. 
Phys. {\bf B433},403 (1995); {\bf B437}, 231 (1995).

\bibitem{fischler} W. Fischler, D. Morgan and J. Polchinski, Phys. Rev. {\bf 
D42}, 4042 (1990).

\bibitem{ff1}The relation 
$M_A=M-\omega$ represents the absorption of negative energy by the black hole, 
associated perhaps with quantum fluctuations.  This is one of the more  
well-known explanations of Hawking radiation \cite{bd}.  Actually, the 
negative energy shell approaches $\hat r=0$;  but this is accompanied by 
an ``outgoing shell" of positive energy Hawking radiation.  

\bibitem{geg}  M. Henneaux, Phys. Rev. Lett. {\bf 54}, 959 (1985);  J. 
Gegenberg and G. Kunstatter, Phys. Rev. {\bf D47},R4192 (1993);  H. Zaidi 
and J. Gegenberg, Phys. Lett. {\bf B328},22 (1994).

\bibitem{bd}  N.D. Birrel and P.C.W. Davies, {\it Quantum Fields in Curved 
Space}, Cambridge, Cambridge Un. Press, 1982.

\bibitem{hh}  J.B. Hartle and S.W. Hawking, Phys. Rev. {\bf D13}, 2188 (1976).

\bibitem{ff2}Figures 1a, 1b, 2a and 
2b are all MAPLE plots.  The units are such that $M=1$.

\bibitem{shell}  For a review of the thin shell formalism, see, {\it
e.g.}, M. Visser, {\it Lorentzian Wormholes}, Chap. 14, New York, AIP
Press, 1995;  see also C. Barrabes and W. Israel, Phys. Rev. {\bf
D43}, 1129 (1991).

\bibitem{ff3}In a recent preprint \cite{micro}, it is claimed that
the back-reaction correction due to Hawking radiation, which, in the
semi-classical regime, consists of replacing $M$ by $M-\omega$ in
expressions for the spectrum, is independent of the details of the
metric.  This is in seeming contradiction to our results, e.g.
\req{3.24}, wherein the back-reaction correction is given by the
replacement of $M$ by $M-11\omega/8$.  We note that there is no
contradiction since the existence of a horizon at some value of $r$
is required by the family of metrics allowed in \cite{micro}.  In
such a case, $\hat R=\hat r$ in \req{geosol}, and we obtain
$\mbox{Im} t=4\pi M_A=4\pi(M-\omega)$, as noted after \req{Imt}.  In
contrast, our allowed class of metrics given by
\req{metricR},\req{metricL} can deviate significantly from a
Schwarzschild geometry near the shell, when the latter is near the
(putative) horizon.  In fact, the geometry bifurcates and the usual
horizon does not exist.  All these effects result from the
reparameterization $r\to R(r)$, which leads to a more general class
of metrics.  For this class, we find that the back reaction is
sensitive to the details of the motion of the shell.

\bibitem{micro} E. Keski-Vakkuri and P. Kraus, ``Microcanonical D-Branes 
and Back Reaction", Cal. Tech. preprint, CALT-68-2079, hep-th/9610045 (1996).

\end{thebibliography}
\end{document}